\documentclass[showpacs,preprintnumbers,twocolumn]{revtex4}
\usepackage{amsmath}
\usepackage{graphicx}
\usepackage{bm}

\input{tcilatex}

\begin{document}

\preprint{}
\title{Detection of Spiral photons in Quantum Optics }
\author{V.V. Klimov$^{\ast }$, D. Bloch$^{\dagger }$, M. Ducloy$^{\dagger }$%
, J.R.Rios Leite$^{\ast }$}
\email{vklim@sci.lebedev.ru;bloch@lpl.univ-paris13.fr}
\affiliation{$^{\ast }$P.N. Lebedev Physical Institute, Russian Academy of Sciences, 53
Leninsky Prospekt, Moscow 119991, Russia;\\
$^{\dagger }$Laboratoire de Physique des Lasers, UMR CNRS 7538 du CNRS et de
l'Universit\U{439} Paris13, 99 Avenue J-B. Cl\U{439}ment F 93430
Villetaneuse, France;\\
$^{\ast \ast }$Universidade Federal de Pernambuco, Centro de Ciencias Exatas
e da Natureza Departamento de Fisica, Cidade Universiteria - Recife -
Pernambuco - Brasil - 50670901}

\begin{abstract}
We show that a new type of photon detector, sensitive to the gradients of
electromagnetic fields, should be a useful tool to characterize the quantum
properties of spatially-dependent optical fields. As a simple detector of
such a kind, we propose using magnetic dipole or electric quadrupole
transitions in atoms or molecules and apply it to the detection of spiral
photons in Laguerre-Gauss (LG) beams. We show that LG beams are not true
hollow beams, due to the presence of magnetic fields and gradients of
electric fields on beam axis. This approach paves the way to an analysis at
the quantum level of the spatial structure and angular momentum properties
of singular light beams.
\end{abstract}

\maketitle

Presently Glauber's theory is widely used to describe the quantum properties
of optical fields and the detection of photons. In [1], he justifies that in
optics, one can restrict in most cases to a detector only sensitive to the
electric field amplitude. The electric dipole (E1) detector he considers, is
assumed to be of negligible size and extra wide frequency band (electric
dipole approximation). Such an assumption will be shown to be restrictive
when the spatial structure of optical fields becomes very complicated. As an
example of such complicated fields one should mention \ complicated optical
fields near nanostructures. Another example of \ optical fields with complex
space structure is the Laguerre-Gauss beams with phase singularity \ and
zero electric field on axis.They are also called spiral beams, and have
attracted a lot of interest owing to their orbital angular momentum[2,3].

The full quantum description of optical fields in such cases is a very
complicated problem, and the development of efficient sensors and photon
detectors in such fields is a very actual task. To characterize such quantum
fields we suggest to use instead of usual E1 detectors, detectors which are
sensitive to gradients of electric fields, or to magnetic fields.

The utility of gradient detectors is already proved in hydroacoustic [4],
where combined receivers, i.e. devices consisting of scalar sound pressure
sensors and several velocity receivers (with mutually perpendicular axes)
are widely used to increase sonar antennas efficiency.

Here, we consider as a click of the detector a specfic atomic excitation (or
the observation of a deexcitation process, such as a fluorescence on a
strong line collected over all space). The specific nature of the detector
is that the atomic exciation is reached through a magnetic dipole or
electric quadrupole transitions

The excitation probability for the most general detector can be found from
usual Fermi's golden rule. In the general case the excitation probability
can be expressed through series of gradients \ of Green function of exciting
quantum field [5]. In the case of coherent narrow band optical fields, one
gets a simpler quasiclassical expression:

\begin{eqnarray}
R_{i\rightarrow f} &=&\frac{\left\vert {T^{if}}\right\vert ^{2}}{\hbar ^{2}%
\sqrt{\delta \omega ^{2}+\Gamma ^{2}/4}}  \label{eq2} \\
T^{if} &=&\mathrm{\mathbf{d}}^{if}\mathrm{\mathbf{E}}\left( {\mathrm{\mathbf{%
r}},\omega _{0}}\right) +\mathrm{\mathbf{m}}^{if}\mathrm{\mathbf{B}}\left( {%
\mathrm{\mathbf{r}},\omega _{0}}\right) +Q_{ij}\nabla _{i}E_{j}\left( {%
\mathrm{\mathbf{r}},\omega _{0}}\right) +...  \notag
\end{eqnarray}%
where $\mathrm{\mathbf{d}}^{if},\mathrm{\mathbf{m}}^{if},Q_{ij}$ \ \ are
matrix elements of electric dipole $\left( E1\right) $, magnetic dipole$%
\left( M1\right) $ and electric quadrupole $\left( E2\right) $transitions
between ground and excited states respectively and where $\delta \omega $
and $\Gamma $ stand for characteristic excitation detuning and transition
linewidth respectively. We also assume that the orientation of the detector
(molecule) is fixed in space and no additional averaging is needed. Although
there can be many Zeeman sublevels in the detector's excited state, only
some of them are of interest for detecting complex optical fields. For
example there can be 3 magnetic dipole ( $m^{M}$, $M=-1,0,1$) and 5
quadrupole transitions ( $Q^{M}$, $M=-2,-1,0,1,2$) the matrix elements of
which can be parameterized in the following form within Cartesian
co-ordinates ( x,y,z) where quantization axis z is chosen along the axis of
the light beam:

\begin{equation*}
\mathbf{m}^{\pm 1}=m^{(1)}\left( \pm 1,i,0\right) ,\mathbf{m}%
^{0}=m^{(0)}\left( 0,0,\sqrt{2}\right)
\end{equation*}

$\mathrm{\mathbf{Q}}^{\left( 0\right) }=Q^{\left( 0\right) }\sqrt{\frac{2}{3}%
}\left\Vert 
\begin{array}{ccc}
-1 & 0 & 0 \\ 
0 & -1 & 0 \\ 
0 & 0 & 2%
\end{array}%
\right\Vert $

$\mathrm{\mathbf{Q}}^{\left( \pm 1\right) }=Q^{\left( 1\right) }\left\Vert 
\begin{array}{ccc}
0 & 0 & \mp 1 \\ 
0 & 0 & -i \\ 
\mp 1 & -i & 0%
\end{array}%
\right\Vert $

\begin{equation}
\mathrm{\mathbf{Q}}^{\left( \pm 2\right) }=Q^{\left( 2\right) }\left\Vert 
\begin{array}{ccc}
1 & \pm i & 0 \\ 
\pm i & -1 & 0 \\ 
0 & 0 & 0%
\end{array}%
\right\Vert  \label{eq2a}
\end{equation}

Usually the main contribution to detector interaction with light is due to $%
\left( E1\right) $ transitions (justifying Glauber's ideal photon detector
[1]). However such transitions give no contribution to excitation of
molecules in regions where there is no electric field! $\left( M1\right) $
and $\left( E2\right) $ transitions give the predominant \ contribution to
excitation rate in this situation. Let us consider this important case in
more details for spiral LG beams.

In the case of LG beams,the electric field can be represented by the
following formulae [2,3]

\begin{equation}
\mathrm{\mathbf{E}}^{m}\left( {\mathrm{\mathbf{r}},\omega }\right) =E_{0}%
\frac{w_{0}}{k}\left\{ {k\alpha U,k\beta U,i\left( {\alpha \frac{\partial U}{%
\partial x}+\beta \frac{\partial U}{\partial y}}\right) }\right\} e^{ikz}
\label{4a}
\end{equation}

with 
\begin{equation}
\begin{array}{l}
U=\frac{C_{p}^{\left\vert m\right\vert }}{w\left( z\right) }\left[ {\frac{%
\sqrt{2}r}{w\left( z\right) }}\right] ^{\left\vert m\right\vert }\exp \left( 
{-\frac{r^{2}}{w^{2}\left( z\right) }}\right) L_{p}^{\left\vert m\right\vert
}\left( {\frac{2r^{2}}{w^{2}\left( z\right) }}\right) \times \\ 
\exp \left( {\frac{ikr^{2}z}{2\left( {z^{2}+z_{R}^{2}}\right) }-im\varphi
-i\left( {2p+\left\vert m\right\vert +1}\right) \arctan \left( {z/z_{R}}%
\right) }\right)%
\end{array}
\label{eq4}
\end{equation}%
where $\ (r,\varphi ,z)$ are cylidrical coordinates, $E_{0}$ is the
amplitude of electric field,$C_{p}^{\left\vert m\right\vert }=\sqrt{2p!/\pi
\left( {p+\left\vert m\right\vert }\right) !}$ is the normalization
constant, $w\left( z\right) =w_{0}\sqrt{1+z^{2}/z_{R}^{2}}$ is the beam
radius at z, $w_{0}$is the Gaussian beam waist, $L_{p}^{\left\vert
m\right\vert }\left( x\right) $ is the generalized Laguerre polynomial, and $%
z_{R}=kw_{0}^{2}/2$ is the Rayleigh range of the beam. $p+1$ gives the
number of nodes of the field in the radial direction.

The most important properties of Laguerre-Gauss beams is that they can carry
both spin and orbital angular momentum, and the total momentum per photon is
given by the formula%
\begin{equation}
j_{z}=\hbar \left( {m+\sigma }\right) ,\text{with \ }\sigma =-i\left( {%
\alpha \beta ^{\ast }-\beta \alpha ^{\ast }}\right)  \label{eq5a}
\end{equation}

where $m\hbar $ is the orbital angular momentum carried by the beam along
its propagation direction [2,3].

The properties of this orbital angular momentum carried by a photon have
been mostly addressed at the macroscopic level, the only experiment
performed to date [6] at the quantum level having attracted much attention
due to the opening of new sets of variable in entanglement. Davila Romero et
al. [7] have recently presented the quantized version of this field. Our
contribution here is to add to this quantum theory the proposal of detectors
for the quanta created in the LG basis.

Below for distinctness we will consider LG beams with $\alpha =1/\sqrt{2}%
,\beta =i/\sqrt{2}$ case. It corresponds to LG beam with spin equal to -1
(circular polarization). So the total angular momentum of our LG beam is
described by $j_{z}=\hbar \left( {m-1}\right) $.

Another key feature of nontrivial $(|m|>1)$ LG beams is the zero of electric
energy density of (\ref{eq4}) on beam axis. Due to this fact LG beams are
often referred to as hollow beams or "doughnut beams" because the electric
field vanishes on the axis.

However we find that magnetic energy density and gradients of electric field
are nonzero at the axis [5]. For example, for LG beams with $m=2$, the
magnetic energy density, $I_{M}$, on the axis is: 
\begin{equation}
I_{M}=\frac{c\left\vert B\right\vert ^{2}}{8\pi }=\frac{cE_{0}^{2}}{8\pi }%
\frac{32\left( {p+1}\right) \left( {p+2}\right) }{\pi \left( {kw_{0}}\right)
^{4}}  \label{eq5}
\end{equation}

\begin{figure}[tbp]
\includegraphics[width=7cm] {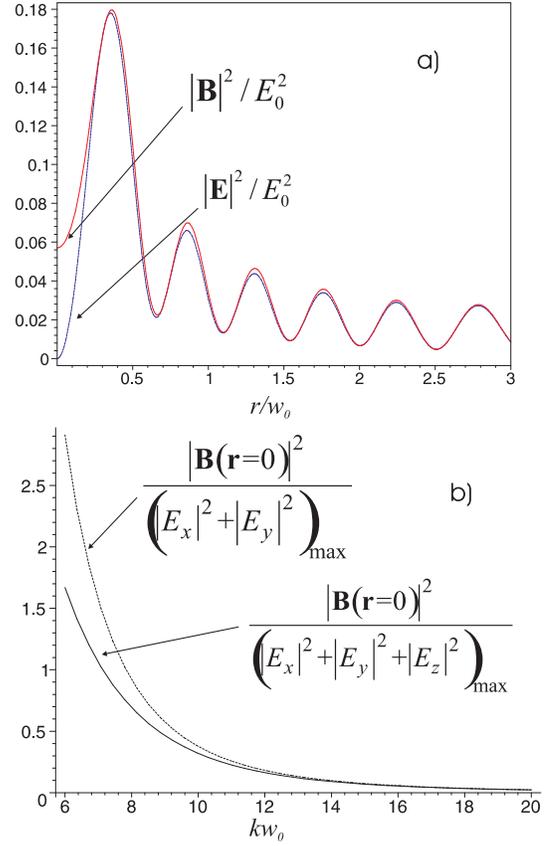}
\caption{(a) Radial dependence of the electric and magnetic energy density
of LG beam in the waist plane $(kw_{0}=10,p=6,m=2)$(b) Ratio of the magnetic
energy density at the center of the LG beam to the electric energy density
at its maximum as function of beam waist $kw_{0}(p=6,m=2)$, as calculated in
the waist plane $z=0$. For a strong focusing, the elementary $(kw_{0})^{-4}$
dependence (dotted line) no longer holds because of a nonnegligible
longitudinal component $E_{z}$ }
\label{fig1}
\end{figure}

Figure \ref{fig1} shows that, for strongly focused beams, the magnetic
energy on axis becomes comparable to the electric energy density at its
maximum. 

The nonzero value of magnetic energy at the beam axis is not an artefact of
the paraxial nature of LG beam [5].It appears for any general nonparaxial
form of monochromatic beam with near cylindrical symmetry [5]

From a formal point of view (Faraday's law of electromagnetic induction),
non zero radial magnetic fields on beam axis are due to the presence of
longitudinal electric fields in the beam. These longitudinal electric fields
are more intense for more focussed beams and for more zeroes in the radial
direction. Deeper insight in this matter shows that nonzero on-axis magnetic
( or electric) fields are related with difficulties in defining a unique
phase for the vector fields.

\begin{figure}[tbp]
\includegraphics[width=7cm] {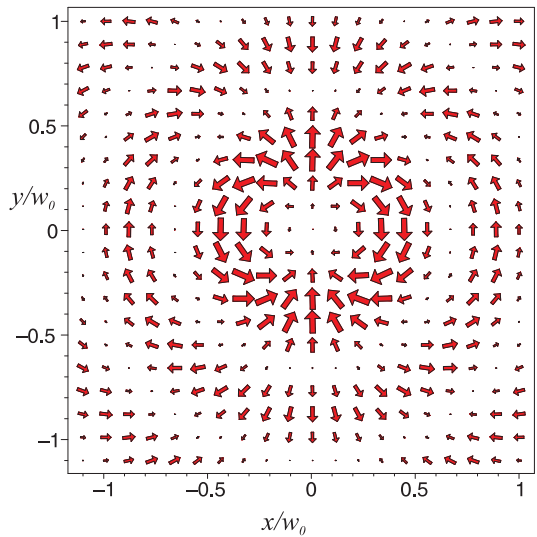}
\caption{Space distribution in the waist plane of the real part of the
electric field in a Laguerre-Gauss beam$%
(kw_{0}=6,p=6,m=2)$. The distribution rotates at the optical frequency }
\label{fig2a}
\end{figure}

\begin{figure}[tbp]
\includegraphics[width=7cm] {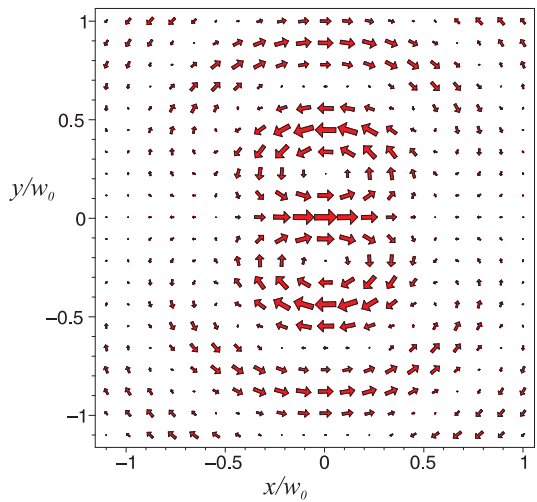}
\caption{Space distribution in the waist plane of the real part of the magnetic field in a Laguerre-Gauss beam$%
(kw_{0}=6,p=6,m=2)$. The distribution rotates at the optical frequency }
\label{fig2b}
\end{figure}

The phase-space structure of Laguerre-Gauss beam is very complicated in
comparison with usual circular polarized light. From Figure \ref{fig2a},\ref{fig2b}
showing the distributions of electric and magnetic field at the waist plane
the complicated magnetic structure of LG beams is evident.


The most interesting feature is that both magnetic field\ (Figure \ref{fig2b}%
) and gradients of electric fields are nonzero at axis so that E1 photon
detector cannot work here. Also even in the case of a linear polarization of
electric fields, magnetic fields will rotate in space with optical frequency.

Obviously very interesting effects can occur with this electromagnetic
energy lying in the region that makes the beam hollow. To test these effects
we suggest using new type of detectors described above. As an example we
have calculated the quadrupole transition amplitudes $(T_{Q}^{mM}=Q_{ij}^{M}%
\nabla _{i}E_{j}^{m}\left( {\mathrm{\mathbf{r}}=0}\right) ,m,M=-2,-1,0,1,2)$
of a molecule placed on the axis of LG beam :

\begin{eqnarray}
T_{Q}^{-1,-2} &=&\frac{4i\sqrt{2p+2}}{\sqrt{\pi }w_{0}}E_{0}Q^{\left(
2\right) }  \TCItag{7a}  \label{eq99a} \\
T_{Q}^{0,-1} &=&\frac{2\left( {8p+4-\left( {kw_{0}}\right) ^{2}}\right) }{%
kw_{0}^{2}\sqrt{\pi }}E_{0}Q^{\left( 1\right) }  \TCItag{7b}  \label{eq100a}
\\
T_{Q}^{1,0} &=&i\frac{4\sqrt{p+1}}{\sqrt{3\pi }}\frac{\left( {8p+8-3\left( {%
kw_{0}}\right) ^{2}}\right) }{k^{2}w_{0}^{3}}E_{0}Q^{\left( 0\right) } 
\TCItag{7c}  \label{eq101aa}
\end{eqnarray}

\begin{equation}
T_{Q}^{2,1}=\frac{8\sqrt{{(p+1)(p+2)}}}{\sqrt{\pi }kw_{0}^{2}}E_{0}Q^{\left(
1\right) }  \tag{7d}  \label{eq102a}
\end{equation}

Other elements of $T_{Q}^{m,M}$vanish.

Eqs. (7a-7d) show that the excitation amplitudes are not symmetric under
(m,M)-$>$(-m,-M) transformation. This is due to interplay between orbital
and spin momenta. As a result, by analyzing (\ref{eq99a}) one can concludes
that such a beam has an angular momentum $j_{z}=\hbar \left( {m-1}\right) $
and $\sigma =-1,$which is in agreement with the independent calculations of
angular momentum of the beam (Eq.\ref{eq5a}) and conservation of angular
momentum. An analogous situation takes place for magnetic$(M1)$ part of
transition amplitudes [5].

\begin{figure}[tbp]
\includegraphics[width=7cm] {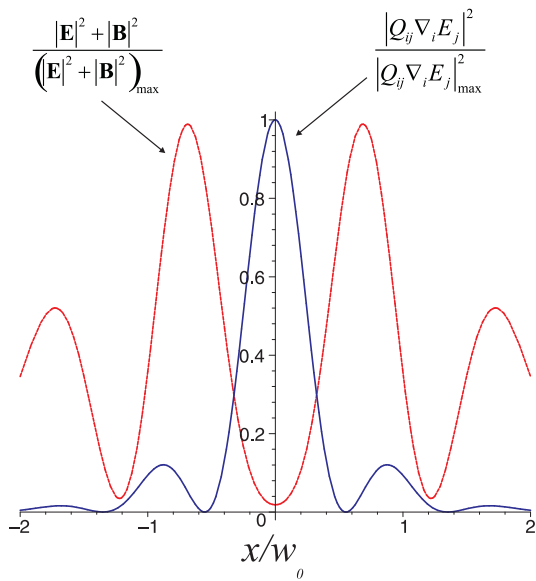}
\caption{Normalized radial distribution of excitation rates of an $E2(M=1)$
transitions of an atom placed in the waist plane of LG beams with\textit{\
p=1,m=2, kw}$_{0}=10.$Dashed line shows normalized distribution of full
energy density. An identical distribution is found for an $M1$ transition .}
\label{fig3}
\end{figure}

The most striking feature is the interaction of $m=2$ LG beam with $M=1$
transitions in molecules. On axis there is no interaction with E1
transition, while there is an efficient interaction with $E2$ and $M1$
transitions. The on-axis excitation rate for such transitions can be written
in the form: 
\begin{equation}
R=E_{0}^{2}\frac{64\left( {p+1}\right) \left( {p+2}\right) }{\hbar ^{2}\sqrt{%
\delta \omega ^{2}+\Gamma ^{2}/4}\pi \left( {kw_{0}}\right) ^{4}}\left\vert {%
m^{\left( 1\right) }+kQ^{\left( 1\right) }}\right\vert ^{2}  \tag{8}
\label{eq8}
\end{equation}%
where $m^{(1)}$ and $Q^{(1)}$ are scalar amplitudes of magnetic dipole and
electric qudrupole transitions (see (\ref{eq2a})).The spatial distribution
of this excitation rates, shown in fig. \ref{fig3} , is also of great
interest. 

One sees that this excitation rate has a well pronounced maximum at the beam
axis. This property should allow exciting selectively molecules located near
the axis, with a sub-wavelength resolution reminiscent of confocal
microscopy. Also, this indicates that a photon can be transfered on-axis, in
spite of the common idea that a LG beam is a hollow beam. It is very
important that a $2\hbar $ exchange of angular momentum is possible in a
single atom-photon transition in contrast to a situation for an $E1$
transition [8]

For such experiments, an issue is certainly the low oscillator strength of
E2 or M1 transitions, usually considered as nearly forbidden transitions.
However, our scheme clearly offers the possibility of detecting processes
for which a Glauber-type detector is blind. Note that the feasibility of
detecting an E2 transition with an evanescent wave (i.e. another type of
e.m. field with a complicated structure [9]) was recently established [10].
In the case of LG beams, a strong focusing, and a large value of $p$, enable
the specific contribution associated to $m=2$ (see eq. 7d) to be comparable
to the common plane-wave contribution (eq. 7b) originating in the
longitudinal field gradient. The lower sensitivity of $E2$ or $M1$
transition, relatively to $E1$ transition used in ideal dipole detector, is
due to the small size of electronic orbit, relatively to optical wavelength
(one has typically $m^{(1)}\sim kQ^{(2)}\sim kea^{2}$, where $a$-
characteristic size of molecules used in detector, while for E1, $%
d^{(1)}\sim ea$). Hence, stronger non-E1 mechanisms are to be expected with
long molecules (e.g. twisted or bio-molecules). A fascinating possibility
apparently offered with LG beams is the selective manipulation of chiral
molecules (the optical activity is usually associated to a coupled $E1-M1$
transition) deposited somewhere close to the hollow region. Although a
negative experimental result was obtained in [11], our investigation of the
local properties of the e.m. field suggests that it is a too large spatial
averaging that has made the effect unobservable.

In conclusion, in this letter, we suggest to use new type of detectors,
which are sensitive to gradients of electric fields and \ to magnetic
fields, in order to detect photons with a complicated space structure. We
have applied our idea to LG beams for which we have shown that spiral beams
should not be considered as hollow, because of nonzero magnetic fields on
the axis of the beam for orbital momentum number $m=2$. Our direct
calculation of excitation rates of such transitions confirm that
Laguerre-Gauss beams bear angular momentum, that can be transferred in an
elementary exchange with a quantum system, hence relaxing the usual
selection rules. Despite we consider coherent state of exciting fields our
conclusions and proposals are valid for any quantum state of exciting field.

From a quantum optics point of view, the investigation of the nature of
spiral photons generated with LG beams has remained until now extremely
limited [6]. Indeed, all experimental investigations involving LG beams and
their specific angular momentum (see [2] and also [6,12,13]) have been
integrated on at least a micron-size volume, instead of using a negligible
size detector. If particle physics considerations had shown that
electromagnetic fields can bear a large angular momentum [14], the coherent
production of large number of identical spiral photons in the optical domain
is a only recent achievement, susceptible to open new frontiers in quantum
optics (e.g. quantum limits to spatial correlation,...). Our suggestion of
using a detector that is not E1-type, but whose size remains intrinsically
microscopic, should help to elucidate experimentally the quantum properties
of these photons carried by a LG or a singular beam, whose specificity
appears enhanced under a strong focusing according to our semiclassical
deriveation. This regime of sharply focused propagating beams also opens a
natural connection with the domain of nano-optics, where it is known that
the relative strength of $E2$ transition is enhanced [15]. More generally,
with the development of nanotechnologies, it becomes conceivable to produce
suitable non-E1 detectors, such as an artificial nanoparticle of special
shape (nanoantennas) designed to be sensitive to gradients of electric
fields. In return, these detectors should benefit to the very contemporary
characterization of more complicated nanooptical fields worth being tested.

VK is grateful to the Russian Foundation for Basic Research (grants
05-02-19647, 07-02-01328) for partial financial support of this work and
University Paris13 for hospitality. VK also thanks V.S.Letokhov for
stimulated discussions. DB MD and RL thanks French Brazilian CAPES-COFECUB
(\#456/04) cooperation support.

\end{document}